\documentclass[a4paper,12pt]{article}
\usepackage[T1]{fontenc}
\usepackage[cp1250]{inputenc}
\usepackage[polish,english]{babel}
\usepackage[cp1250]{inputenc}
\usepackage{graphicx}
\usepackage[numbers]{natbib}

\title{Comments on evolution of cooperation: 
evolutionary stability in enhanced Dove-Hawk model}
\author{Pawel Sobkowicz\footnote{e-mail address: {\tt pawelsob@poczta.onet.pl}}}
\date{11th October 2003}

\begin{document}

\maketitle

\section{Abstract}

One of the best examples of  traditional analysis of evolutionary stable strategies (ESS) is provided by the so called Dove-Hawk model. In this paper we present several enhancements to the model aimed at describing the evolution of cooperative behavior. In addition to Doves and Hawks we introduce several groups of Cooperators, who act as Doves within their own group, but as Hawks outside it. This allows to study how cooperating groups may grow and achieve stability within a nonrepeating evolutionary games framework.   
Depending on initial conditions, the final stable population may have one, all-encompassing Cooperator population or several competing cliques. After taking into account that Cooperators bear costs necessary to recognize members of one's own group it is possible to see populations where Cooperators eventually lose against Hawks or populations where several cliques of Cooperators coexist with a Hawk population. 

\section{Theoretical background}
The classical works by Maynard Smith\cite{maynard73-1,maynard82-1} introducing and popularizing the evolutionary game theory and the concept of Evolutionary Stable Strategy have resulted in widespread interest in such modeling approach by social scientists, biologists and economists. 

Within the evolutionary games theory one describes a population consisting of $N_{TOT}$ `players', who interact among themselves. The form of the interaction is simplified to `contests', which are single events during which two players decide how to share among themselves some valuable resource. For simplicity it is usually assumed that the resource value is constant (denoted by $V$). For `peacefully' resolved contests the value is divided among the participant players (equally or unequally). In situations where contests develops into a fight over the resource, players participating in the fight may suffer injuries, expend unnecessary resources etc., which is modeled by including a penalty cost $-h$ (harm). For simplicity the cost $h$ is assumed to be equal for both players participating in a fight.

There are two main variants of evolutionary games: Nonrepetitive, single instance contests, in which participants have no memory of previous encounters and thus cannot adapt their strategies to past behaviour of their opponents, and Repetitive Games, where the  memory allows complicated strategies, individual behavior and learning. 

Within this paper we concentrate on Nonrepetitive Games. Even for such simplified situation  players may use different strategies. To achieve the status of Evolutionary Stable Strategy, specific behaviour $\sigma$ must either: do better against itself than any other variant strategy or, if some mutant strategy would do  just as well against $\sigma$ as $\sigma$ itself, then $\sigma$ must do better against the mutant than the mutant against itself \cite{maynard73-1}.

The modeling off the population dynamics is best performed assuming that the activities take place within separate rounds or iterations. Such iterations may correspond, for example, to yearly cycles or biological generations -- or may not correspond to any discernible quantization of the modeled system, but be a convenient simplification. Within such iterations players interact among themselves and compete for resources. Between iterations, following the average outcomes (payoffs) for different groups, the composition of the total player population is adjusted. After several (many) iterations one can  determine which of the strategies succeeds and becomes an ESS.

The best known of such models is the Hawk-Dove model of Maynard Smith and Price\cite{maynard73-1}. The population consists of two groups: Doves ($D$) and  Hawks ($H$).  These groups are best  characterized by their behavior in contests. When a Dove meets a Dove, they resolve the contest peacefully, without a fight. The value $V$ is then, statistically, split in half between the contestants. Thus the average outcome (payoff) for a player in a D-D match is $V/2$. When a Dove meets a Hawk, the latter `bullies' the Dove and gets the whole value of $V$, leaving the Dove empty handed. On the other hand, when two Hawks meet a fight ensues and the average payoff is $V/2 - h$. For $V/2-h > 0$ (if only `pure' strategies are allowed) the Hawk becomes an ESS, by virtue of winning the contests with the Doves.

The Hawks and Doves model serves as perfect example that an evolutionary stable strategy is not necessarily the one that returns thee best results for the participating players. In a pure Dove population the average payoff of $V/2$ is greater than in Hawk population. Nevertheless, because the outcome of inter-`species' encounters are always favorable to Hawks, even a small Hawk population would invade a Dove society and have evolutionary advantage. In a Dove population invaded by Hawks, the average payoff diminishes as the proportion of Hawks grows.

The aim of this paper is to describe some consequences of the extension of the Hawk-Dove model within a nonrepetitive paradigm. The extension is based upon an introduction of a new strategy, called here `Cooperator'. A Cooperator behaves like a Dove --- when it competes with a member of its own group. But when a Cooperator encounters a  Hawk or a Dove it behaves as a Hawk. To obtain more general results we introduce several separate Cooperator groups, denoted by $C_i$, which act friendly within their own group, but competitively outside it. In the simplest case, with just one Cooperator group, the dynamics of the system is trivial. The Cooperator payoff is always better than the payoff for a Hawk. Cooperators could easily invade a Dove society. Thus single group Cooperator strategy is an ESS. 

It is worth noting, that although we play the game without `memory', to allow the Cooperator-like behavior some sort of recognition of group membership is necessary. This can be achieved through some labeling, pre-contest behaviour etc\footnote{As always where labeling is present, it opens the field for mimicry or cheating behavior. Effects of such strategies will be analysed elsewhere}.

As we are interested in exploring further further than the single Cooperator group, that is into the formation and stability of populations with several competitive Cooperator groups  $C_i$. Although in general one expects Cooperators to fare better than Hawks and Doves, it is interesting to see in what conditions stable population compositions involve just one dominant clique and when more cliques form and persist. We have enhanced the range of their possible behaviours by including the possibility of transfer of Cooperators from different groups (recruitment). In particular, we propose a model in which during an encounter between two cooperators from different groups first check  possibility of recruitment (that is one of the players may consider joining the other's group `\textit{on the spot}') --- and then both players behave friendly in the encounter. Only when recruitment fails the two players resort to competitive strategy. In general, inclusion of recruitment complicates the model, but allows  modeling of phenomena such as social group dynamics. One should remark, that such recruitment is possible when Cooperators not only invest in some means of recognizing agents from various groups (including one's own), but also have some means of establishing the reasons for joining (such as results in previous iteration, or group size) during an encounter.

To give our model more precise form, let's define $N_D$ as number of Doves, $N_H$ as number of Hawks, $N_{C_i}$ as number of Cooperators of cooperator group $C_i$. The number of Cooperator groups is denoted by $N_{CG}$. We assume that the population size is constant, $N_{TOT}$
\begin{equation}
N_{TOT} = N_D + N_H + \sum_i N_{C_i}
\end{equation} 
Additionally we define $N_C = \sum_i N_{C_i} $.

For a given iteration we introduce average payoffs for each of the groups, $P_D$, $P_H$ and $P_{C_i}$. Assuming that simulation round consists of contests of all players among themselves we easily derive:
\begin{equation}
P_D = \frac{(N_D-1)V/2}{N_{TOT}} 
\end{equation} 
\begin{equation}
P_H = \frac{N_D V + (N_H-1) (V/2-h) + N_C (V/2-h)}{N_{TOT}} 
\end{equation} 
\begin{eqnarray}
\label{pci} 
P_{C_i} & = & \frac{N_D V + (N_H) (V/2-h) + (N_{C_i}-1) V}{N_{TOT}} \nonumber \\ 
 & + & \frac{\sum_{j \ne i} \left( p_{j\rightarrow i} N_{C_j}  V/2 +
(1-p_{j\rightarrow i}) N_{C_j} (V/2-h) \right) }{N_{TOT}} 
\end{eqnarray} 

Within equation \ref{pci} the probability of player from group $j$ joining the group of just encountered player $i$ is denoted by $p_{j\rightarrow i}$. Such a probability may, of course be modelled in many ways. For the purposes of this paper we use the following assumptions:
\begin{eqnarray}
p_{j\rightarrow i} & = & {0 \mbox{ for } \delta_{ji} < T_j = t \tilde{P}_{C_j}} \nonumber \\
p_{j\rightarrow i} & = & {f(\delta_{ji})  \mbox{ for } \delta_{ji} \ge T_j = t \tilde{P}_{C_j}}
\end{eqnarray} 
where $\delta_{ji}$ is the difference in payoffs for Cooperator group $i$ and group $j$ in the previous iteration:
\begin{equation}
\delta_{ji} = \tilde{P}_{C_i} - \tilde{P}_{C_j}
\end{equation}

The presented model reflects two aspects of the recruitment: a player would agree to join a group with better results than his own ($\delta_{ji} > 0$) and it would join only if the perceived benefits (based on previous round results) exceeded certain threshold $T_j$. We have set a simple model of the threshold being a fixed proportion ($t$) of the results  achieved by player's own group in previous round.

When the threshold
 is exceeded, probability of recruitment becomes nonzero. Different theoretical approaches may use different forms describing the recruitment probability. We have decided to model the probability by relatively simple form of function $f$:
\begin{equation}
f(\delta_{ji}) = \frac{\delta_{ji}-T_j}{\delta_{ji}+T_j}
\end{equation}
With this choice probability is small for payoff differences just above the threshold, while they grow to $\sim 1$ when the payoff difference is considerable (compared to the threshold for each group), see Figure~1.

So far the `intra-iteration' population dynamics were discussed. When the phase describing agents encounters is finished, then, according to the values of payoffs for each `species' and group we model the `inter-iteration' dynamics. Here again, flexibility of choice for the mechanisms is considerable. Within this work we use a relatively straightforward form of updates to populations.

The payoff for each group $P_{\alpha}$  (where $\alpha$ stands for $D$, $H$, or $C_i$) is compared to the average payoff $P_{AV}$ for the whole population. 
Normalized deviation for group $\alpha$ is defined as:
\begin{equation}
D_{\alpha} = P_{\alpha}/P_{AV} -1.
\end{equation}
We define then
\begin{equation}
D = \sqrt{\sum_{\alpha} D_{\alpha}^2}
\end{equation}  
\begin{equation}
\phi_{\alpha} = D_{\alpha}/D.
\end{equation} 
The new size of group $N_{\alpha}'$ is then modelled as follows:
\begin{eqnarray}
N_{\alpha}'  =  \frac{N_{\alpha} e^{\phi_{\alpha}/S}}{M} & \hbox{ if }  &|D_{\alpha}| > T_G \nonumber \\
N_{\alpha}'  =  N_{\alpha} & \hbox{ if } & |D_{\alpha}| \leq  T_G 
\end{eqnarray}
where $M$ is a normalizing factor ensuring that the total population remains constant and $N_{\alpha}$ is the size of the group after the recruitment phase. $S$ is an artificial damping factor used to decrease the scale of inter-iteration changes and smooth out simulations. Effect of $S$ is marginal -- it affects mostly the number of simulations until stability is reached, but not the final outcome. The threshold $T_G$ is introduced to model the situation in which relatively small (i.e. smaller than $T_G$) differences in payoffs should not have significant effect on population dynamics. The reasoning behind such a threshold is based on observation that in `real life' situations, factors other than direct competitiveness influence the survival and changes in population and may smooth out the payoff differences if these are sufficiently small.

\section{Recognition costs}

One of the most interesting simple enhancements is an inclusion of `labeling cost', that is the cost borne by Cooperators to provide reliable means of recognition and determination of group membership. The costs may have different origin and form, depending on the modelled system (such as behavioral costs, costs of group uniforms etc.). 

We propose here a simple approach in which the labeling cost $l$ is equal for  all Cooperators and is applied to {\bf all} contests, {\bf before} the contest is resolved. This corresponds to Cooperators investing in specific behavior or a recognizable `uniform'   displayed at all meetings with the purpose of showing the group membership. In such case the payoff for Cooperators in all contests is diminished by $l$. Thus the expected payoff for a Cooperator from group $C_i$ becomes:

\begin{eqnarray}
\label{pcil} 
P_{l,C_i} & = & \frac{N_D (V -l)+ (N_H) (V/2-h-l) + (N_{C_i}-1) (V-l)}{N_{TOT}} \nonumber \\ 
 & + & \frac{\sum_{j \ne i} \left( p_{j\rightarrow i} N_{C_j}  (V/2-l) +
(1-p_{j\rightarrow i}) N_{C_j} (V/2-h-l) \right) }{N_{TOT}} 
\end{eqnarray}   

Obviously for $l>h$ Cooperators are at distinct disadvantage to Hawks and do not have any chance to become an ESS. On the other hand, for the opposite situation ($l<h$) we have quite interesting situations. Lets first concentrate on a simple case with just one group of Cooperators, competing with Hawks and Doves. 

The difference in expected payoff for the Cooperators and Hawks becomes
\begin{eqnarray}
P_{l,C}-P_H & = &\frac{1}{N_{TOT}} \left( N_D (V-l) + (N_C-1) (V/2-l) + N_H (V/2-h-l) \right. \nonumber \\
 & &\left.  -N_D V - N_C (V/2-h) - (N_H-1) (V/2-h) \right) 
\end{eqnarray}   
assuming that $N_H, N_C \gg 1$ we can simplify the above expression to
\begin{equation}
P_{l,C}-P_H = \frac{1}{N_{TOT}} \left( N_C(h-l) - (N_H+N_D) l  \right) 
\end{equation} 
As the Doves are at disadvantage and after sufficiently long evolution would die out, we have a simple condition 
\begin{equation}
P_{l,C} > P_H  \Longleftrightarrow \frac{N_C}{N_H} > \frac{l}{h-l}
\end{equation}.
The interpretation of the above condition is that a small cooperator group would not be able to invade a well established Hawk society. Thus the ratio $l/(h-l)$ determines the course of evolution through initial conditions (as discussed in the next section).

\section{Results of simulations}

As the number of possible combinations of the input parameters is enormous, we have selected several examples of simulation results, showing typical behavior. These results are presented in the following Tables 1 and 2 and Figures 2-9. In the presented simulation examples we have started from relatively large Dove population (92\% of the total population) and small admixtures of Hawks and Cooperators. The  initial sizes od the Cooperator groups were assigned randomly. Payoff parameters were assigned in a way to reflect non prohibitive fighting costs and labeling costs ($V>h>l$).

\begin{table}[h]
\begin{small}
\begin{tabular}{|l|c|c|c|c|c|c|c|c|} \hline
\multicolumn{9}{c}{Initial populations } \\ \hline \hline
{\sl Simulation} & A & B & C & D & E & F & G & H\\ \hline
$N_{TOT}$ & 10000 & 10000 & 10000 & 10000 & 10000 & 10000 & 10000 & 10000\\ \hline 
$N_{D}$ & 9200 & 9200 & 9200 & 9200 & 9200 & 9200 & 9200 & 9200\\ \hline 
$N_{H}$ & 500 & 300 & 300 & 300 & 300 & 300 & 150 & 150\\ \hline 
$N_{CG}$ & 6 & 6 & 6 & 6 & 6 & 6 & 6 & 6\\ \hline 
\multicolumn{9}{c}{Payoff and threshold parameters  } \\ \hline \hline
$V$     & 12  & 12  & 12  & 12   & 12   & 12  & 12  & 12\\ \hline 
$h$     & 4   & 4   & 4   & 4    & 4    & 4   & 4   & 4\\ \hline 
$l$     &  0  & 0   & 0   & 0    & 0    & 0   & 0   & 0\\ \hline 
$t$     & 0.1 & 0.2 & 0.2 & 0.25 & 0.25 & 0.3 & 0.3 & 0.1\\ \hline 
$T_G$   & 0.1 & 0.1 & 0.2 & 0.25 & 0.20 & 0.2 & 0.3 & 0.1\\ \hline 
\multicolumn{9}{c}{Final (stable) populations } \\ \hline \hline
$N_{D}$ & 0   & 0   & 0   & 0    & 0    & 0   & 0   & 0\\ \hline 
$N_{H}$ & 0   & 0   &   0 &   0  &   0  &   0 &   0 & 0\\ \hline 
$N_{CG}$& 1   & 1   & 1   & 2    & 1    & 2   & 4   & 2\\ \hline 
$N_{C_i}$&10000&10000&10000&5392 &10000 &6133 & 3435& 5431\\ 
         & 0  & 0   & 0   & 4608 & 0    &3861 & 3278& 4569\\ 
         & 0  & 0   & 0   & 0    & 0    & 0   & 1675& 0\\
         & 0  & 0   & 0   & 0    & 0    & 0   & 1607& 0\\
         & 0  & 0   & 0   & 0    & 0    & 0   & 0   & 0\\
         & 0  & 0   & 0   & 0    & 0    & 0   & 0   & 0\\    \hline
\end{tabular} 
\end{small}
\caption{Examples of simulation results for labeling cost $l=0$. Columns A—H describe  different simulation runs.}
\end{table} 

\begin{table}[h]
\begin{small}
\begin{tabular}{|l|c|c|c|c|c|c|c|c|} \hline
\multicolumn{9}{c}{Initial populations } \\ \hline \hline
{\sl Simulation} & I & J & K & L & M & N & O & P\\ \hline
$N_{TOT}$ & 10000 & 10000 & 10000 & 10000 & 10000 & 10000 & 10000 & 10000\\ \hline 
$N_{D}$ & 9200 & 9200 & 9200 & 9200 & 9200 & 9200 & 9200 & 9200\\ \hline 
$N_{H}$ & 300 & 300 & 300 & 150 & 170 & 300 & 300 & 300\\ \hline 
$N_{CG}$ & 6 & 6 & 6 & 6 & 6 & 6 & 6 & 6\\ \hline 
\multicolumn{9}{c}{Payoff and threshold parameters  } \\ \hline \hline
$V$     & 12  & 12  & 12  & 12   & 12   & 12  & 12  & 12\\ \hline 
$h$     & 4   & 4   & 4   & 4    & 4    & 4   & 4   & 4\\ \hline 
$l$     &  1  & 1   & 1.5 & 1.5  & 1.5  & 2   & 2   & 2\\ \hline 
$t$     & 0.1 &0.25 & 0.2 & 0.1  & 0.2  & 0.1 & 0.3 & 0.5\\ \hline 
$T_G$   & 0.1 &0.25 & 0.2 & 0.1  & 0.2  & 0.1 & 0.3 & 0.5\\ \hline 
\multicolumn{9}{c}{Final (stable) populations } \\ \hline \hline
$N_{D}$ & 0   & 0   & 0   & 0    & 0    & 0   & 0   & 5184\\ \hline 
$N_{H}$ & 0   & 5654&10000&   0  & 6077 &10000&   0 & 4343\\ \hline 
$N_{CG}$& 1   & 6   & 0   & 1    & 2    & 0   & 0   & 6\\ \hline 
$N_{C_i}$&10000&1301& 0   &10000 & 2060 & 0   & 0   & 118\\ 
         & 0  & 1357& 0   & 0    & 1863 & 0 & 0   & 116\\ 
         & 0  & 865 & 0   & 0    & 0    & 0   & 0   & 92\\
         & 0  & 796 & 0   & 0    & 0    & 0   & 0   & 86\\
         & 0  & 13  & 0   & 0    & 0    & 0   & 0   & 43\\
         & 0  & 13  & 0   & 0    & 0    & 0   & 0   & 17\\    \hline
\end{tabular} 
\end{small}
\caption{Examples of simulation results for nonzero labeling cost $l$.}
\end{table} 

A few general trends can be seen from the presented simulation results.
First, for all cases except P, population of Doves decreases to zero. This is an obvious result of the disadvantages that Doves have when faced with more belligerent groups. In the presented simulation in which  Doves do not die out, we have assumed relatively high value of $T_G$ threshold, equal to 50\% of the average payoff. For such large value, after some iterations, the populations become frozen (see Figures 8 and 9), as the payoff difference is too small to affect any changes within our model.

Second, for simulations with zero labeling constant and `reasonably' small thresholds, Hawks die out as well. Obviously, the ability to resolve contest costlessly within a Cooperator group always gives  Cooperators a better position. 

Interesting behavior may be seen when the labeling cost is nonzero. Then, depending on the initial set of control parameters the end result may be, in addition to Cooperators, Hawk population may exist. The dynamics of the population becomes very dependent not only on the control parameters, but also on the initial conditions. For example, simulations K and M have the same set of control parameters, differing only in the initial number of Hawks and Cooperators. Although in both cases, the respective sizes of the populations $N_H$ and $N_C$ are smaller then few percent of the total population size, the final composition is radically different. This illustrates the fact that the initial conditions can play significant role  in the population dynamics.

\section{Discussion}

\subsection{Hawk-Dove-Possessor}
In a recent paper \cite{yee02-1}, Yee has presented analysis of the evolutionary stability of a different example of enhancement of the Hawk-Dove game. The analysis was aimed at situations where there are definite resources that can be `kept' between iterations  or claimed \emph{before} contests. Yee described results of a strategy called `Possessor' (P), which is defined as follows: Possessor behaves like a Hawk if it is currently the `owner' of the resource (defending its right to is vigorously) and like a Dove if it is not an owner (accepting the `first come first served' mentality). The new strategy, even though conceptually very simple proves to be effective against Hawks and Doves. Moreover, behavior like this is observed both in human societies and in animals. Nevertheless, the Possessor strategy, does not lead to cooperation between players, and is actually a very individualistic position.

\subsection{Crucial role of thresholds to obtain nontrivial solutions}
In contrast to the repeated PD games and other studies including learning, encounter memory and adaptive strategies of players the model presented in this paper does not touch the issue of \emph{formation} of Cooperator like behaviour and Cooperator groups. Our aim was rather to study the factors which might influence the dynamics of the populations, especially whether it is possible to obtain \emph{stable and mixed} populations.

The general tendency shown by simulated populations shows that, indeed, such mixed populations are possible. The key role allowing their formation is played by the two thresholds introduced into the model. Without those thresholds arbitrarily small differences in payoffs may dominate the population evolution. As mentioned before, the commonsense reason for the introduction of the thresholds has its roots in the cumulative uncertainty of the effects of the  factors which may determine the payoff for individual player, which are not modelled in our work. Such `noise' (for example luck of individual players in finding necessary resources) would wash out the payoff differences between groups resulting from discussed strategies if they are sufficiently small. 

There is an interesting similarity between the presented model and the analysis of Frank \cite{frank88-1} of the need of investment in the display and recognition of intentions and threats. Such displays are important in repetitive PD games, ensuring cooperative behaviour. In our case, within the framework of Nonrepetitive PD games, labeling allows recognition of group membership and thus of the strategy of the agents in an encounter before the costly competition takes place. To recognize the group membership (and thus the strategy) the agents must, however, pay `recognition costs'.  Such situation leads to stable mixed population, because the benefit for Cooperators at being able to recognize Defectors and shirk from meeting with could be balanced by the recognition costs.

The key result of the paper is that a very simple model may lead to the establishment of several types of populations --- corresponding to situations found in biology or sociology. Depending on initial conditions, the final stable population may have one, all-encompassing Cooperator population or several competing cliques. After taking into account labeling costs it is possible to simulate  populations where Cooperators eventually lose against Hawks or populations where several cliques of Cooperators coexist with a Hawk population.


\clearpage

\section{Figures}

\begin{figure}[h]
\centering
\includegraphics[height=10cm]{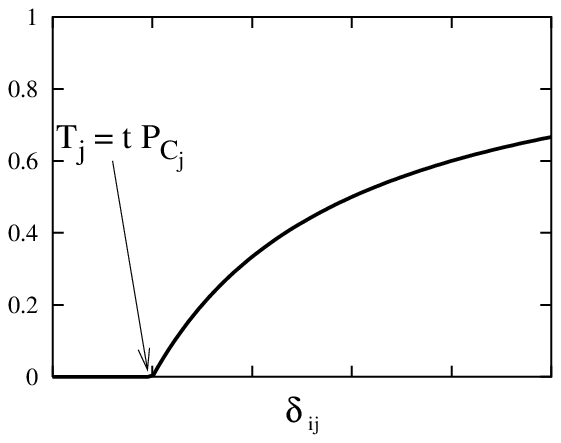}
\caption{Probability of recruitment $p_{j\rightarrow i}$ as function of the payoff difference $\delta_{ji}$. }
\end{figure}

\begin{figure}[h]
\centering
\includegraphics[height=10cm]{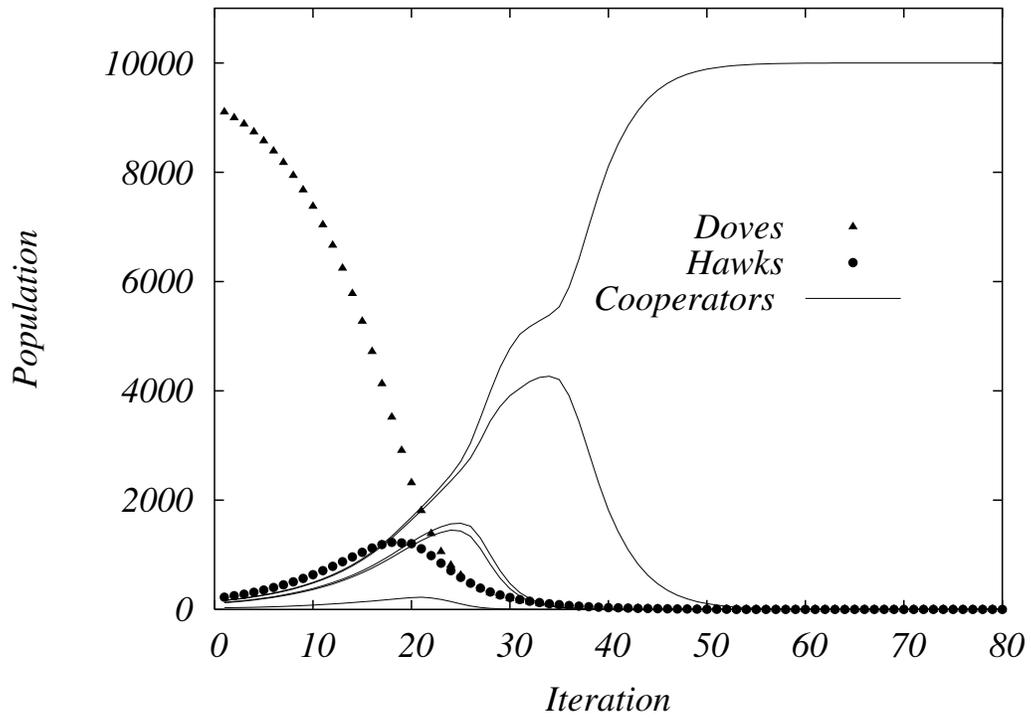}
\caption{Example 1 --- Evolution of population composition ending in one dominant Cooperator group. Parameters used: $V=12$, $h=4$, $l=0$, $t=0.1$, $T_G=0.1$. }
\end{figure} 
\begin{figure}[h]
\centering
\includegraphics[height=10cm]{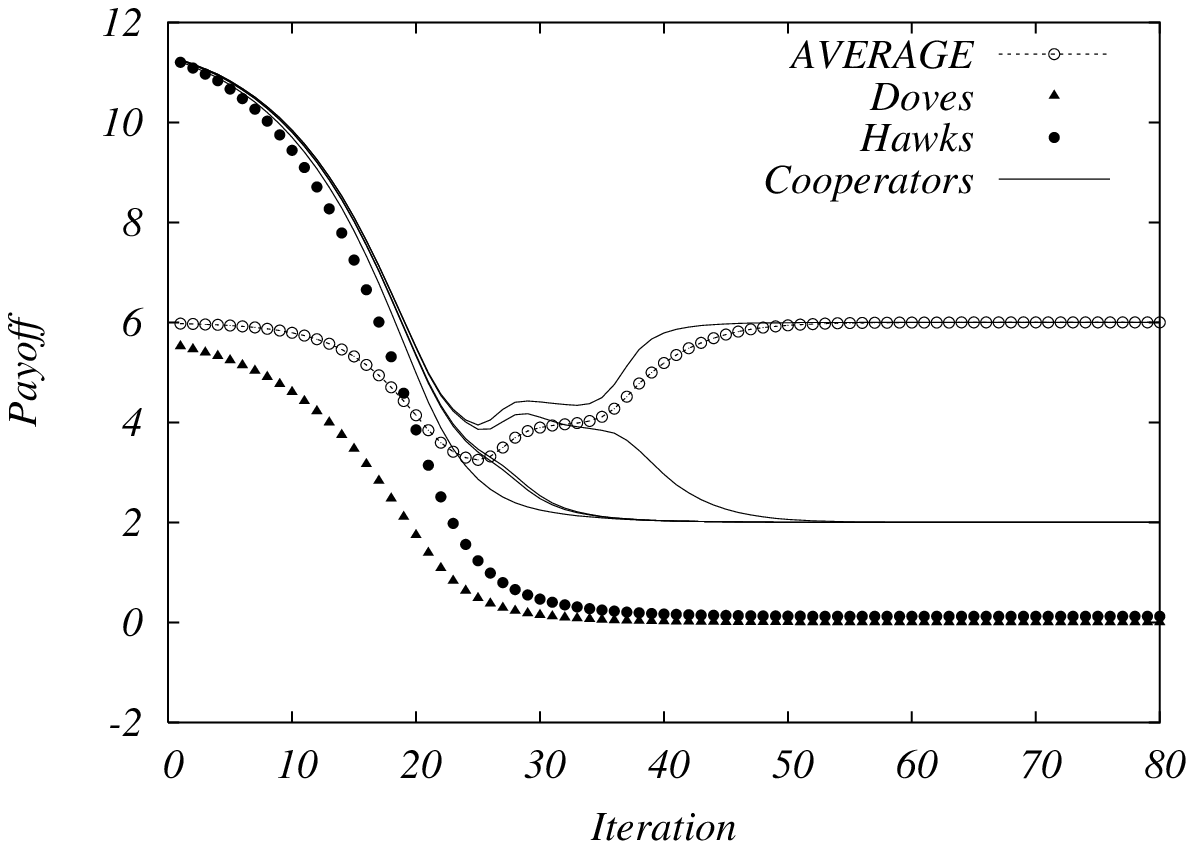}
\caption{Evolution of payoffs for the Example 1}
\end{figure} 
\begin{figure}[h]
\centering
\includegraphics[height=10cm]{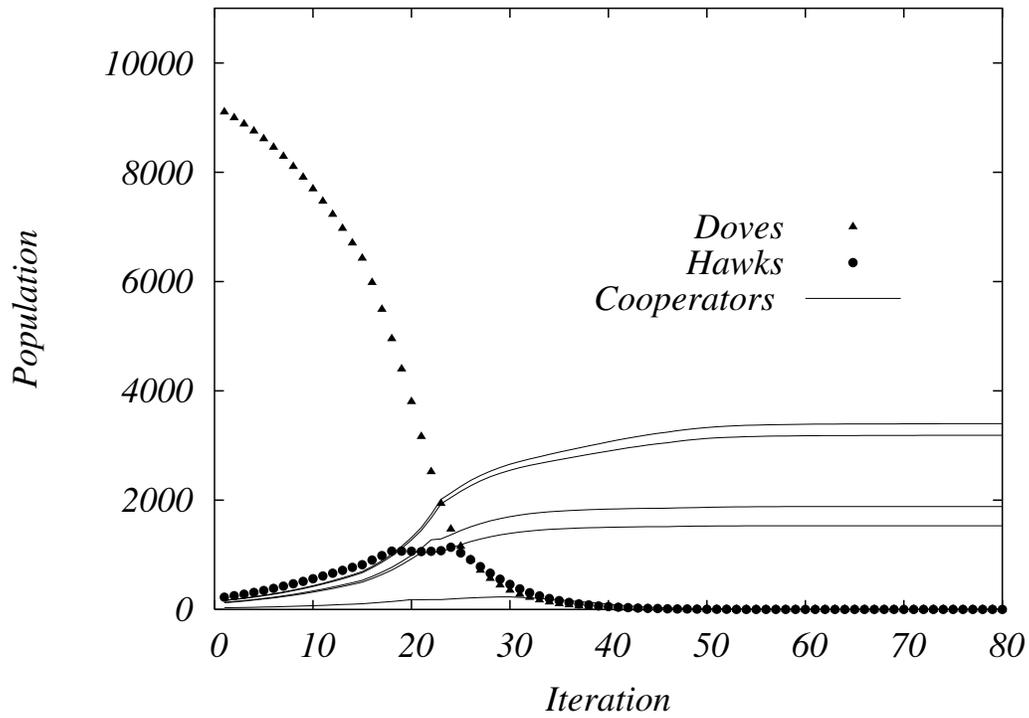}
\caption{Example 2 --- Evolution of population composition ending in several Cooperator groups. Parameters used: $V=12$, $h=4$, $l=0$, $t=0.3$, $T_G=0.3$. }
\end{figure} 
\begin{figure}[h]
\centering
\includegraphics[height=10cm]{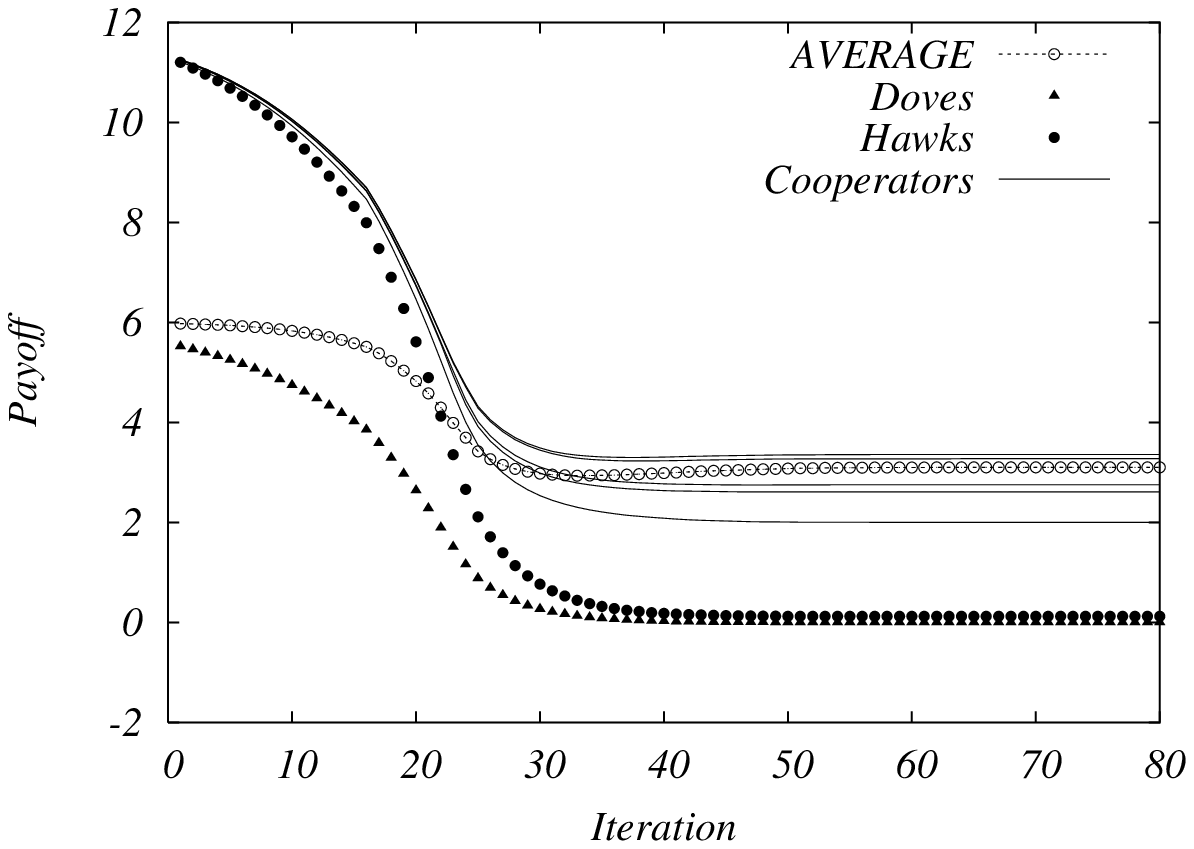}
\caption{Evolution of payoffs for the Example 2}
\end{figure} 
\begin{figure}[h]
\centering
\includegraphics[height=10cm]{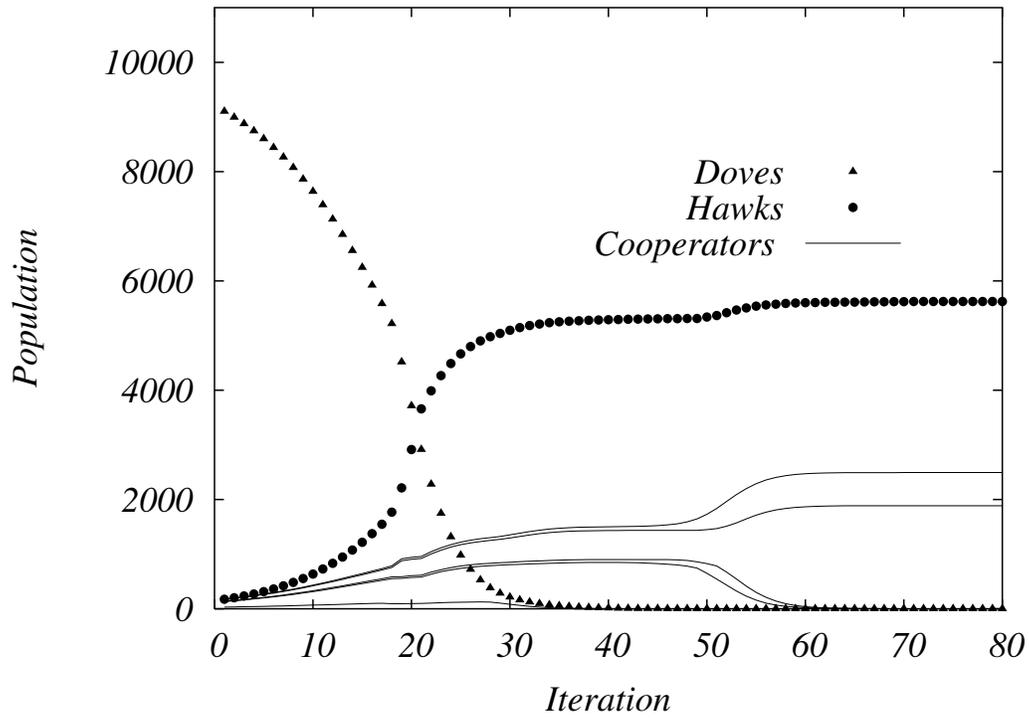}
\caption{Example 3 --- Evolution of population composition for simulation with nonzero labeling cost ending in several Cooperator groups and Hawk population. Parameters used: $V=12$, $h=4$, $l=1.5$, $t=0.3$, $T_G=0.3$. }
\end{figure} 
\begin{figure}[h]
\centering
\includegraphics[height=10cm]{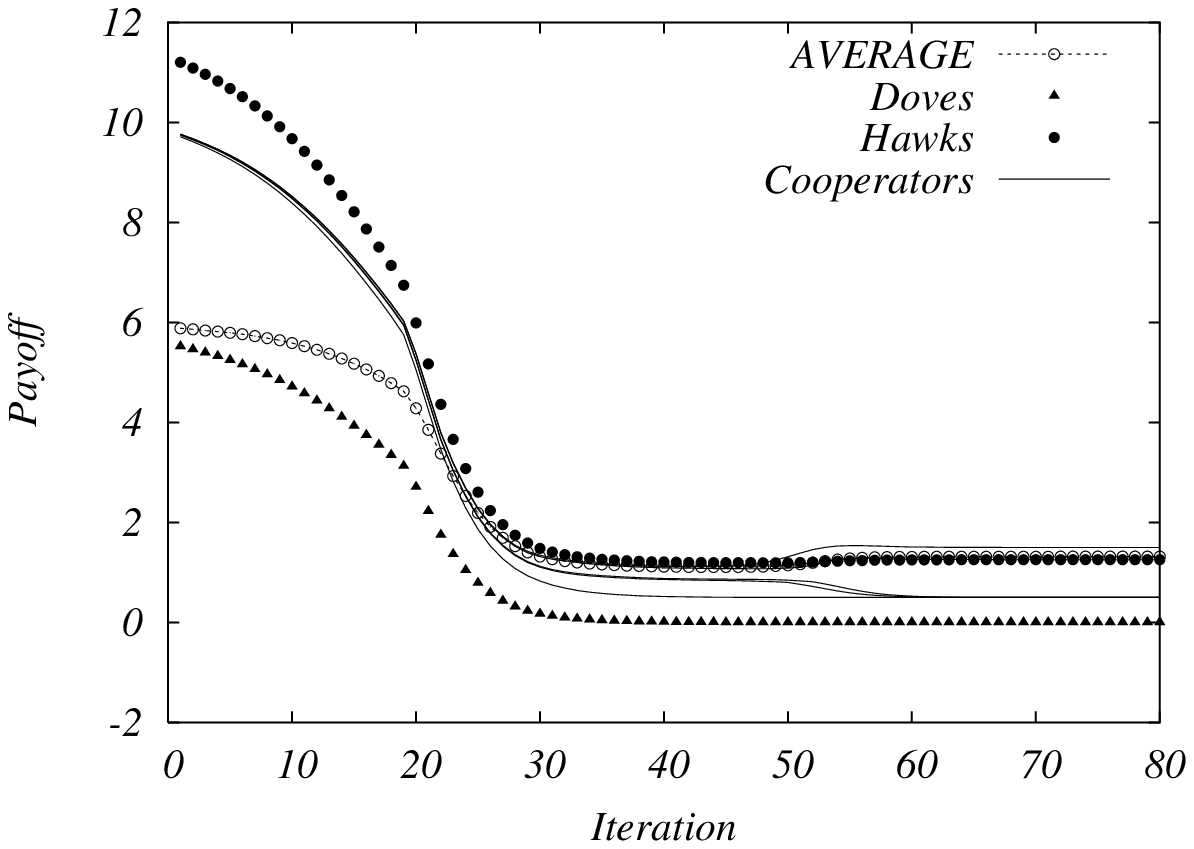}
\caption{Evolution of payoffs for the Example 3}
\end{figure} 

\begin{figure}[h]
\centering
\includegraphics[height=10cm]{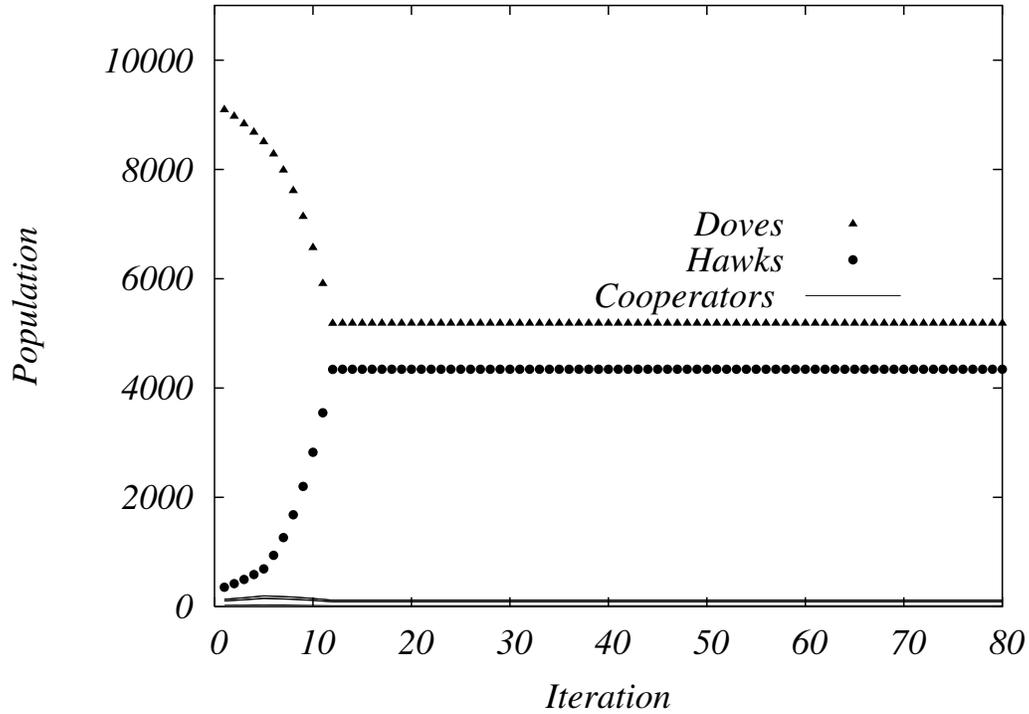}
\caption{Example 4 --- Evolution of population composition for simulation with nonzero labeling cost ending in mixed Dove, Hawk and several Cooperator groups. Due to quite high $T_G$ value of $0.5$ after several iterations differences in payoffs no longer change thee populations. Parameters used: $V=12$, $h=4$, $l=2$, $t=0.5$, $T_G=0.5$. }
\end{figure} 
\begin{figure}[h]
\centering
\includegraphics[height=10cm]{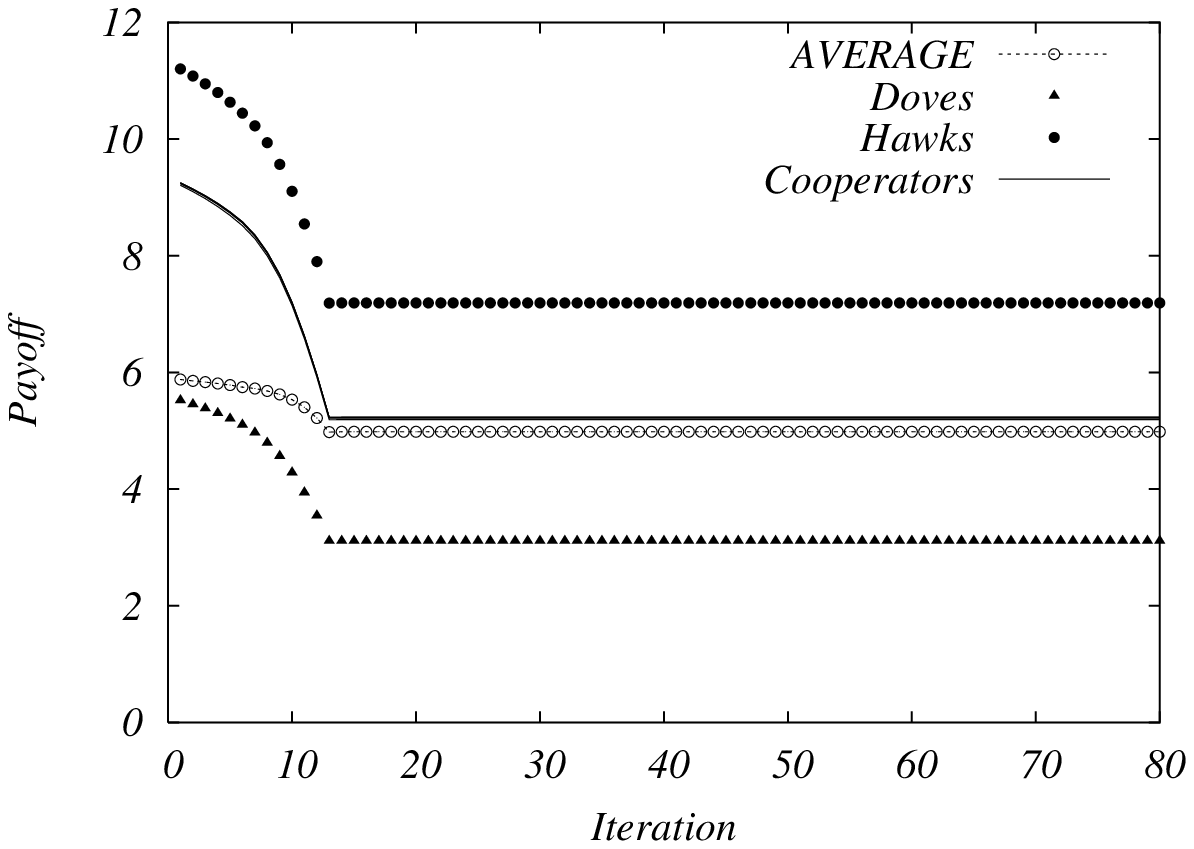}
\caption{Evolution of payoffs for the Example 4}
\end{figure}

 \end{document}